\documentclass[11pt,a4paper]{article}
\usepackage{jheppub_my}
\usepackage{amssymb}
\usepackage{graphicx}
\usepackage{amsmath}

\graphicspath{{Fig_reduced/}}

\newcommand{\bS}{\ensuremath{\mathbb{S}}}
\newcommand{\bW}{\ensuremath{\mathbb{W}}}
\newcommand{\bZ}{\ensuremath{\mathbb{Z}}}

\newcommand{\scH}{\ensuremath{\mathcal{H}}}
\newcommand{\scI}{\ensuremath{\mathcal{I}}}
\newcommand{\scN}{\ensuremath{\mathcal{N}}}
\newcommand{\scW}{\ensuremath{\mathcal{W}}}

\newcommand{\matW}[8]{\ensuremath{
W
\left(\begin{array}{cc}
#8 & #7 \\
#5 & #6 \\ 
\end{array}
\right)
}}

\newcommand{\beq}{\begin{equation}\begin{aligned}}
\newcommand{\eeq}{\end{aligned}\end{equation}}

\newcommand{\mmod}[1]{[\! [#1]\! ]}

\title{New Integrable Models \\ from the Gauge/YBE Correspondence}
\author{Masahito Yamazaki$^{\blacklozenge, \lozenge}$}

\preprint{IPMU-13-0134}

\affiliation{$^{\blacklozenge}$Kavli Institute for the Physics and Mathematics of the
Universe (WPI), \\ the University of Tokyo,  Chiba 277-8583, Japan}

\affiliation{$^{\lozenge}$Princeton Center for Theoretical Science, Princeton University, Princeton, NJ 08544, USA}

\abstract{
We introduce a class of new integrable lattice models
labeled by a pair of positive integers $N$ and $r$.
The integrable model is obtained from
the Gauge/YBE correspondence,
which states the equivalence of 
the 4d $\scN=1$ $S^1\times S^3/\bZ_r$ index of a large class of 
$SU(N)$ quiver gauge
theories with the partition function of 
2d classical integrable spin models.
The integrability of the model (star-star relation) is 
equivalent with the invariance of the index under 
the Seiberg duality.
Our solution to the Yang-Baxter equation 
is one of the most general known in the literature,
and reproduces a number of known integrable models.
Our analysis identifies the Yang-Baxter equation
with a particular duality (called the 
Yang-Baxter duality) between two
4d $\scN=1$ supersymmetric quiver gauge theories.
This suggests that the integrability goes beyond
4d lens indices and can be extended to the 
full physical equivalence among the IR fixed points.
}

\begin{document}

\maketitle

\section{Introduction}\label{sec.intro}

The goal of this paper is to introduce a class of new integrable models
with the help of exact results in supersymmetric gauge theories.
The integrability here refers to solutions of the Yang-Baxter equation
(YBE). 

Our integrable model will be constructed from the following relation between
4d $\scN=1$ quiver gauge theories and the 2d integrable models,
which we call the {\it Gauge/YBE correspondence}:
\beq
\scI_\textrm{4d index}[S^1\times S^3/\bZ_r]= Z_\textrm{2d spin}[C] \ .
\label{GaugeYBE}
\eeq

On the left hand side of \eqref{GaugeYBE}
we have the 4d lens index \cite{Benini:2011nc}
for a class of 4d $\scN=1$ supersymmetric $SU(N)$ quiver gauge theories
defined from quiver diagrams on a 
two-dimensional geometry $C$, which is either
a disc \cite{Xie:2012mr,Franco:2012mm}
or a torus \cite{Hanany:1997tb,Hanany:1998it,Hanany:2005ve}.
The 4d lens index
is a twisted partition function on $S^1\times S^3/\bZ_r$
and is a function of several fugacities.
Two of the fugacities will be denoted by $p,q$,
and others for global symmetries will be parametrized by
$R_i$.

One the right hand side of \eqref{GaugeYBE}
we have the 
partition function of a 2d classical integrable spin system
defined on $C$, where the quiver diagram is identified with the lattice of the spin chain.
The boundary condition of the spin chain is either periodic or fixed 
depending on whether $C$ is a torus or a disc.
The parameters of the index, $p,q$ and $R_i$, are translated into the 
temperature-like parameters and the 
spectral parameters of the spin chain, respectively.

The correspondence \eqref{GaugeYBE}
holds for each value of $r$ and $N$.
The $r=1$ case of \eqref{GaugeYBE} was discussed in the 
earlier paper by the author
and the collaborators \cite{Yamazaki:2012cp,Terashima:2012cx,Xie:2012mr}.
There the 4d $S^1\times S^3$ index \cite{Kinney:2005ej,Romelsberger:2005eg} 
of the 4d gauge theories
was identified with the 
statistical mechanical integrable model discussed
in  \cite{Bazhanov:2010kz,Bazhanov:2011mz,Bazhanov:2013bh}
(see also \cite{Spiridonov:2010em}).

For the more general case $r>1$ discussed in this paper,
we can use the correspondence \eqref{GaugeYBE} to {\it define}
a new statistical model from the known gauge answer on the left hand side,
for each pair of positive integers $(r,N)$.
The spin at a vertex of the lattice has both discrete and
continuous spins, each of which has $N$ components with $1$ constraint imposed. 
The R-matrix is written in terms of lens elliptic gamma functions
(Appendix \ref{sec.gamma}),
and is a function of
two temperature-like parameters $p,q$ as well as the spectral parameters
$R_i$. 
Our solution to YBE is one of the most general known in the literature,
see more comments in section \ref{sec.comments}.

The origin of the integrability of the 2d spin system
has a clear-cut explanation 
on the
supersymmetric gauge theory side of the 
correspondence \eqref{GaugeYBE}:
the star-star relation \cite{Baxter:1986,Bazhanov:1992jqa} is precisely the 
invariance of the 4d lens index 
under Seiberg duality (see \cite{Spiridonov:2010em} for the case with $r=1$).
The star-star relation is stronger than the YBE and implies the latter.

We will identify the counterpart of YBE
to be a duality (called the Yang-Baxter duality)
between two 4d $\scN=1$
quiver gauge theories, which can be obtained by composition/gauging of four
$\scN=1$ Seiberg dualities.
Since this is a physical duality, 
we expect that the discussion of integrability here 
actually goes beyond the 4d lens indices discussed in this paper,
see comments in section \ref{sec.conclusion}.

The identification of integrable models and the 4d $\scN=1$
gauge theories in the Gauge/YBE correspondence is summarized in Table \ref{tab.summary}.

\begin{table}[htbp]
\caption{Dictionary in the Gauge/YBE correspondence.}
\begin{center}
{\renewcommand\arraystretch{1.2}
\begin{tabular}{c|c}
integrable model & 4d $\scN=1$ gauge theory \\
\hline
spin lattice & quiver diagram \\
rapidity line & global symmetry \\
rapidity (spectral) parameter & global symmetry fugacities \\
partition function & 4d lens index (on $S^1\times S^3/\bZ_r$) \\
temperature-like parameters & fugacity $p, q$ \\
$U(1)$ spin variables & Wilson line along thermal $S^1$ \\
$\bZ_r$ spin variables & Wilson line along the Hopf fiber \\
number of spin components & rank of a gauge group \\
self-interaction & $\scN=1$ vector multiplet \\
nearest-neighbor interaction & $\scN=1$ bifundamental chiral multiplet  \\
star-star relation & $\scN=1$ Seiberg duality \\
R-matrix & theory $\mathcal{T}[R]$ (Figure \ref{fig.faceweight}) \\
composition of R-matrices & gauging \\
Yang-Baxter equation & Yang-Baxter duality \eqref{catYBE} \\
high-temperature expansion & dimensional reduction
\end{tabular}
}
\end{center}
\label{tab.summary}
\end{table}

\bigskip
The rest of this paper is organized as follows.
In section \ref{sec.model} we describe our statistical mechanical model,
whose gauge theory origin is explained in section \ref{sec.quiver}.
Further comments on the model are included in section
\ref{sec.comments},
and we conclude with summary and 
comments on gauge theory implications in 
section \ref{sec.conclusion}.
In Appendix \ref{sec.gamma}
we summarize the lens elliptic gamma function used in the paper.

We hope that the paper is readable both for 
experts on supersymmetric gauge theories 
and integrable models.
Experts on integrable models will
find the complete definition of the statistical 
mechanical model in section \ref{sec.model},
without referring to the gauge theory explanation in 
section \ref{sec.quiver}.
Experts on supersymmetric gauge theories 
will benefit from section 
\ref{sec.quiver}, which elucidates otherwise ad-hoc
definitions in section \ref{sec.model}.

\section{New Integrable Models}\label{sec.model}

In this section we spell out the definitions of our integrable lattice
spin models.
Our construction follows closely the $r=1$ case of
\cite{Bazhanov:2011mz} (see also \cite{Baxter:1997tn}).\footnote{However 
some details are changed from \cite{Bazhanov:2011mz}
for the better match with supersymmetric gauge theories.
For example, the direction of 
parallel rapidity lines (to be introduced momentarily)
are all the same in \cite{Bazhanov:2011mz}, whereas
they are alternating here. Our choice of rapidity parameters is 
slightly more general than in \cite{Bazhanov:2011mz}.
}

\subsection{Definition of the Model}

\paragraph{Spin Lattice}

\begin{figure}[htbp]
\centering{\includegraphics[scale=0.3]{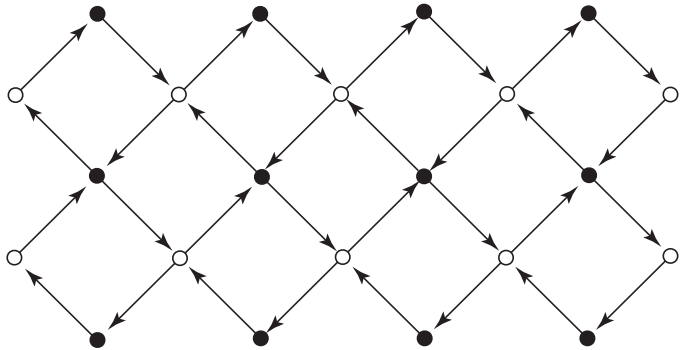}}
\caption{The spin lattice of the statistical mechanical model.
The spins reside at the vertices of the lattice, and their
 nearest-neighbor interactions are represented by edges. 
The interaction is chiral, which is represented by the orientation of
 the edges. The color of a vertex (black or white) is determined by the orientations
of arrows around it.}
\label{fig.lattice}
\end{figure}

Let us consider a periodic spin lattice shown in Figure \ref{fig.lattice}.
We denote the set of vertices and edges of the lattice by $V$ and $E$.
Note that the edges in the lattice are oriented:
for an edge $e$ we denote its head (target) and tail (source) by $h(e), t(e)\in V$.
Note also that around each vertex the orientations of the arrows
alternate, and hence the number of incoming and outgoing arrows are the
same.
We color a vertex either black or white 
depending on the orientations of the arrows around the vertex.
The lattice is then bipartite, {\it i.e.}\ an edge always connect
two vertices of different colors.\footnote{This bipartite graph is
{\it not} the bipartite graph discussed in
\cite{Yamazaki:2012cp,Terashima:2012cx};
it is the spin lattice (which will be identified with a
quiver diagram) 
which is bipartite here.}
In this section we consider the square lattice,
see however comments on generalization towards the end of section \ref{sec.quiver}.

\paragraph{Spin Variables}

We place spin variables 
at each lattice site $v\in V$:
continuous $U(1)$ $N$-component spins 
$z_{v,1}, \ldots, z_{v,N} \in e^{2\pi\sqrt{-1} \mathbb{R}} \, (|z_{v,i}|=1)$
 and discrete $N$-component spins $m_{v,1}, \ldots, m_{v,N} \in \bZ_r$.
We moreover impose the condition that 
\beq
\prod_i  z_{v,i}=1 \ , \qquad \sum_i m_{v,i}\equiv 0 \quad (\textrm{mod
} \, r) \ ,
\label{zprod}
\eeq 
and hence only $N-1$ of $z_{v,i}$ ($m_{v,i}$) are independent.
In the following we collectively denote the spins at a vertex $v$
by
$s_v=(z_v, m_v)$.

Note that we here consider classical (not quantum) statistical mechanical models,
and spin here just refers to dynamical variables of the theory over which we sum/integrate over.
On the boundary of the lattice we can impose periodic boundary conditions
or fixed boundary conditions. In the latter case the boundary spins will be non-dynamical
and the partition function will be a function of the values of the boundary spins.
For simplicity we in the following will mostly consider periodic
boundary conditions: the case with other boundary condition
is similar, but some part of the analysis requires careful analysis of boundary
effects.

\paragraph{Rapidity Parameters and Rapidity Lines}

For the purpose of discussing integrability let us introduce extra parameters to the 
statistical model.
These parameters $R_e$ are associated for edges $e\in E$,
and we impose that condition that
\beq
\sum_{e: \textrm{ around a vertex } v}  (1-R_e)=2 \ , \quad 
\sum_{e: \textrm{ around a face }}  R_e =2 \ .
\label{sumR}
\eeq
It is these parameters $R_e$ which enter into the definition of the
partition function. 

\begin{figure}[htbp]
\centering{\includegraphics[scale=0.3]{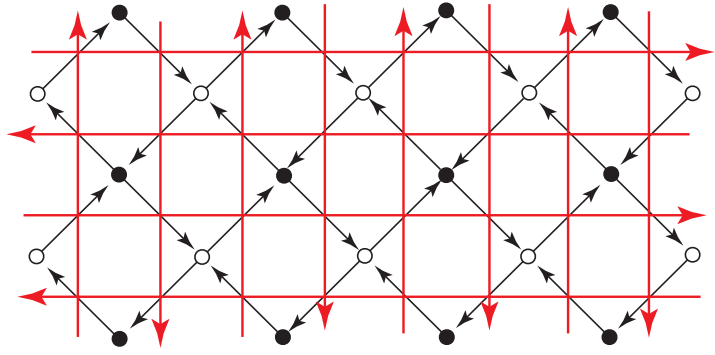}}
\caption{The rapidity lines for the spin lattice 
of Figure \ref{fig.lattice}, shown in red arrows.
}
\label{fig.latticewithlines}
\end{figure}

\begin{figure}[htbp]
\centering{\includegraphics[scale=0.3]{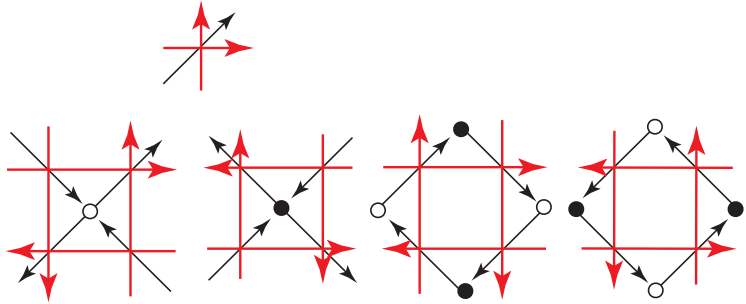}}
\caption{We can reconstruct the spin lattice from the rapidity lines, 
following the rule shown here.}
\label{fig.latticerule}
\end{figure}

The conditions \eqref{sumR} do not have unique solutions.
However the ambiguity can be naturally
parametrized by the rapidity lines
\cite{Baxter:1978xr}\footnote{Rapidity 
lines are called zig-zag paths in the context of
dimer models, which are the dual graphs of the spin lattice 
discussed in this paper.}, which are drawn as 
red line in Figure \ref{fig.latticewithlines}.
Rapidity lines contain the same data as the original 
spin lattice --- we can recover the original lattice from the set of rapidity
lines, following the rules in Figure \ref{fig.latticerule}. 

\begin{figure}[htbp]
\centering{\includegraphics[scale=0.3]{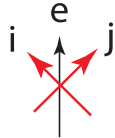}}
\caption{The rapidity parameter for an edge $e$ is given by the
 difference of $R_i$ and $R_j$ associated with two rapidity lines $i, j$.}
\label{fig.rapidityrule}
\end{figure}

Let us denote the set of rapidity lines by $I$,
and its element by $i\in I$. We associate a rapidity parameter
(spectral parameter)
$R_i$ to each rapidity line $i$.
Given a solution $R_e$ to \eqref{sumR}
we can modify $R_e$ to be\footnote{Since we have a
difference of two rapidities here
we have the so-called rapidity difference
property.} 
\beq
R_e':=R_e+R_i-R_j \ ,
\label{difference}
\eeq
when two rapidity lines $i, j$ pass through the edge $e$
as in Figure \ref{fig.rapidityrule}:
We can then easily verify that $R_e'$ also satisfy \eqref{sumR}.
Note that the overall shift of $R_i$
do not change the values of $R_e$, and 
hence is irrelevant for the statistical mechanical model of this paper.

\paragraph{Boltzmann Weight}

An edge $e\in E$, starting from a vertex $t(e)$ and ending on another $h(e)$, 
represents the nearest-neighbor interaction
between the spins $s_{t(e)}$ and $s_{h(e)}$. 
The corresponding Boltzmann weight is given by
\begin{equation}
\begin{split}
\bW^e(R;s)&=\bW^e_{R_e}(s_{t(e)}, s_{h(e)})\\
&=\mathcal{W}^e_0(s_{t(e)}, s_{h(e)})\, \prod_{1\le i,j\le N} \,
\Gamma_{r, \mmod{m_{t(e),i}-m_{h(e),j}}}\left(
(pq)^{\frac{R_e}{2}} \frac{z_{t(e),i}}{z_{h(e),j}}
;p,q\right) \ ,
\end{split}
\label{Wweight}
\end{equation}
where $\Gamma_{r,\mmod{m}}$ is the lens elliptic gamma function 
and $\mmod{m}$ denotes $m$ modulo $r$ (see Appendix \ref{sec.gamma}).
The multiplicative factor $\mathcal{W}^e_0(s_{t(e)}, s_{h(e)})$ is given by
\begin{equation}
\begin{split}
 \mathcal{W}^e_0(s_{t(e)}, s_{h(e)})=&
 \prod_{1\le i,j \le N} \Bigg[ (p\, q)^{\frac{1}{4 r}\mmod{m_{t(e),i}-m_{h(e),j} } \mmod{-m_{t(e),i}+m_{h(e),j} } 
(1-R_e) } \\
& \times
\, \left(\frac{p}{q} \right)^{-\frac{1}{12 r} \mmod{m_{t(e),i}-m_{h(e),j} } \mmod{-m_{t(e),i}+m_{h(e),j}}
(2 \mmod{m_{t(e),i}-m_{h(e),j} }-r)} \\
& \times  \left(\frac{z_{t(e),i}}{z_{h(e),j}} \right)^{-\frac{1}{2 r} \mmod{m_{t(e),i}-m_{h(e),j} } \mmod{-m_{t(e),i}+m_{h(e),j} }} 
\Bigg]
\ .
\end{split}
\end{equation}
Note that this Boltzmann weight is chiral ({\it i.e.}\ not reflection
symmetric) $\bW^e_{R_e}(s_{t(e)}, s_{h(e)})\ne \bW^e_{R_e}(s_{h(e)},
s_{t(e)})$, except for the special case $N=2$.

We also include self-interactions among the 
different components of the spin $s_v$ at the same vertex $v$.
The Boltzmann weight for the self-interaction at a vertex $v\in V$
is given by
\beq
\bS^v(s)=\bS^v(s_v)
=
\left( \prod_{a=0}^{r-1} \frac{1}{n_{v,a}!} 
\right)
\mathcal{S}^v_0(s_v)
\left( \Gamma_{r, 0}(1;p,q) \right)^{-(N-1)} 
\prod_{i \ne j }\, \Gamma_{r, \mmod{m_i-m_j}}\left(\frac{z_{v,i}}{z_{v,j}};p,q\right)^{-1} 
\ ,
\label{Sweight}
\eeq
where
we denoted the number of 
$a \in \{0,1, \ldots, r-1\}$ in $\{m_{v,1}, \ldots, m_{v,N} \}\in
(\bZ_r)^N$ to be $n_{v, a}$:
by definition we have $\sum_{a=0}^{k-1} n_{v,a}=N$.
The multiplicative factor $\mathcal{S}^v_0(s_v)$, which depends only on the 
discrete spins $m_v$ out of $s_v$, is given by
\beq
\mathcal{S}^v_0(s_v)= (p \, q)^{-\frac{1}{4 r}\sum_{i\ne j} \mmod{m_{v,i}-m_{v,j} } \mmod{-m_{v,i}+m_{v,j} } } \ .
\eeq
Note that the parameters $R_e$ do not appear in $\bS^v$.

It is often possible to absorb this contribution into the 
definition of $\bW^e(s_{t(e)}, s_{h(e)})$, however
it will be more natural to keep this factor for the identification with gauge theories.

\paragraph{Partition Function}

The partition function is defined to be the statistical sum over all (continuous as well as 
discrete) spin configurations $\{s_v \}_{v\in V}$:
\begin{equation}
\begin{split}
Z
&=\sum_{s_v\, (v\in V)} \,
\, 
\left( \prod_{v\in V}  \bS^v(s)\right) 
\left( \prod_{e\in E} \bW^e(R;s) \right) \\
&=\sum_{m_v\, (v\in V)} \int_{|z_{v,m}|=1}  \prod_{v\in V}
 \prod_{i=1}^{N-1}  \frac{d z_{v,i}}{2\pi \sqrt{-1} z_{v,i}}\ 
\left( \prod_{v\in V}  \bS^v(z,m)\right) 
\left( \prod_{e\in E} \bW^e(R;z,m) \right) \ ,
\label{Ztotal}
\end{split}
\end{equation}
where in the integrand we write $z_{v,N}$ in terms of $z_{v,i}$ $(i=1, \ldots, N-1)$ 
in terms of \eqref{zprod}. 

The total partition
function \eqref{Ztotal} depends on 
two temperature-like parameters $p, q$ as well as rapidity parameters $R_i$.
All of these parameters will appear inside the Boltzmann weight of the
model. We can either regard $p, q$ as formal parameters, or impose $|p|, |q|<1$ for convergence of the Boltzmann
weights.

\paragraph{Reformulation as IRF Model}

The statistical mechanical model defined here 
can be reformulated 
either as a vertex model or 
an IRF (interaction-round-a-face) model \cite{Bazhanov:2011mz}.
To avoid repetition we
in the following
use the language of IRF models.

\begin{figure}[htbp]
\centering{\includegraphics[scale=0.25]{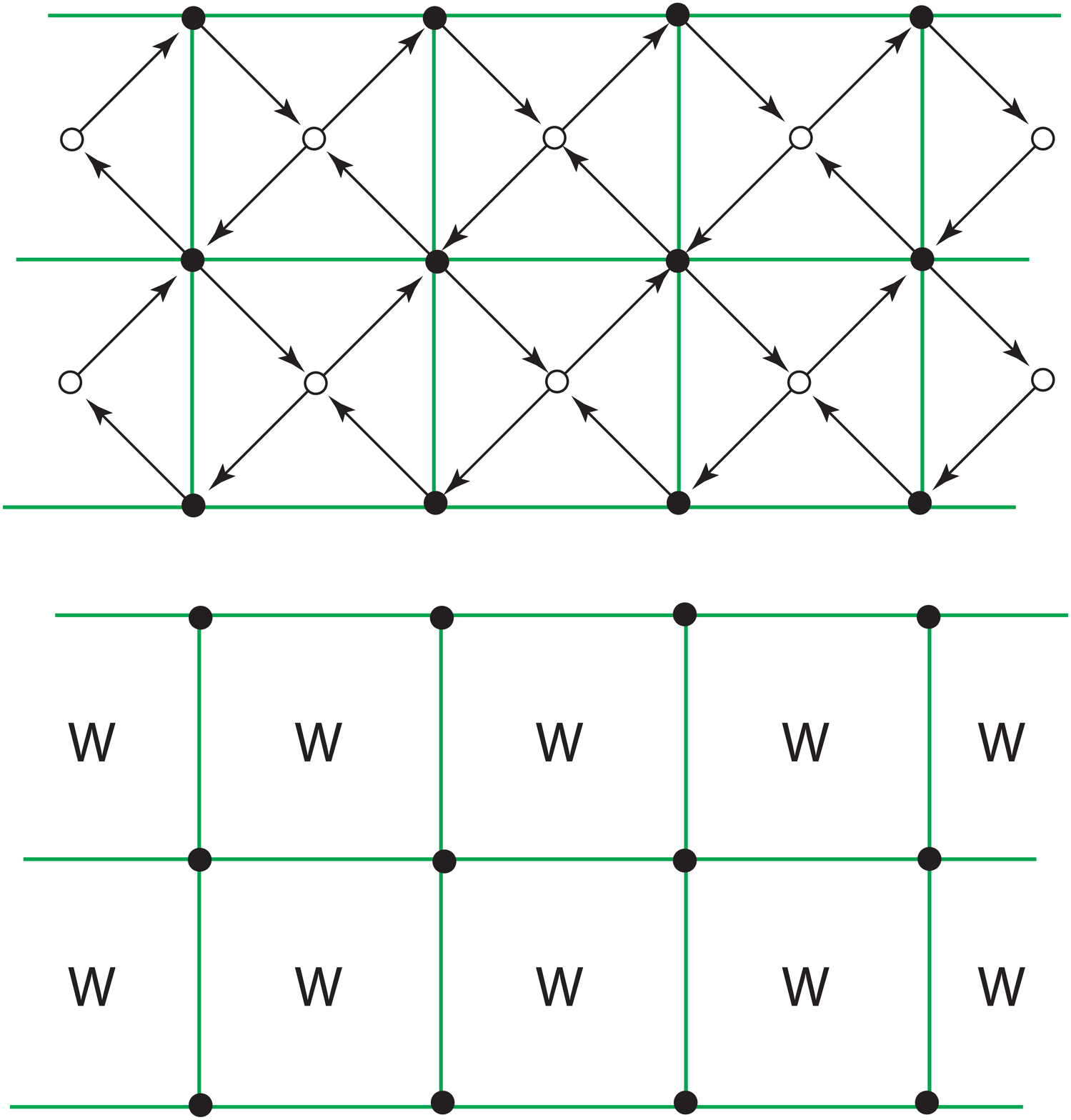}}
\caption{The reformulation as an IRF model. The spins at white
vertices are now integrated out, leaving only the black vertices
defined on the green lattice. The Boltzmann weight $W$ is
then associated with the faces of the new lattice.
Alternative possibility is to exchange the role of white and black
 vertices,
leading to another IRF model defined on the dotted lattice.
The two IRF models nevertheless have the same partition function.
}
\label{fig.IRF}
\end{figure}

To define an IRF model, we simply 
integrate out spin variables 
at white vertices,
while keeping those spins at black vertices.
The remaining lattice is then a larger lattice 
colored green in 
Figure \ref{fig.IRF},
and the Boltzmann weights are associated with the faces of the 
new lattice. 

\begin{figure}[htbp]
\centering{\includegraphics[scale=0.3]{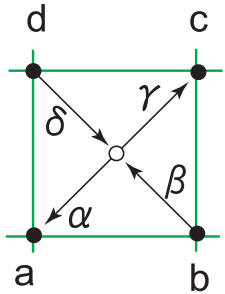}}
\caption{The face weight for the IRF model is determined
from the Boltzmann weights for four edges, as well
as those of four vertices, of the original spin lattice.
We have denoted the four black vertices by $a, b, c, d$, 
and the weights for the edges connected by them by 
$\alpha, \beta, \gamma, \delta$. The first of the constraint equations \eqref{sumR}
in this case means $\alpha+\beta+\gamma+\delta=2$.
}
\label{fig.faceweight}
\end{figure}

Suppose that we have a face $F$,
with vertices $a, b, c, d$ as in Figure
\ref{fig.faceweight}. We denote the rapidity parameters of 
four edges by $\alpha, \beta, \gamma, \delta$.
It follows from \eqref{sumR} that they satisfy the relation
$\alpha+\beta+\gamma+\delta=2$.
We take the Boltzmann weight for the face $F$
to be
\begin{equation}
\begin{split}
W(F)&=W\left(
\begin{array}{cc}
 d & c \\
 a & b \\
\end{array}
\right) \\
&= 
\sqrt{\frac{\bW_{2-\gamma-\delta}(d,c) \bW_{2-\beta-\gamma}(b,c)}
{\bW_{2-\alpha-\beta}(a,b)\bW_{2-\alpha-\delta}(a,d)}}
\sqrt{\bS^a \bS^c}\, \sum_g 
\bS^g \, \bW_{\alpha}(a,g) \bW_{\beta}(g,b) \bW_{\gamma}(c,g)
\bW_{\delta}(g,d) \ ,
\end{split}
\label{Wdef}
\end{equation}
where we used the short-handed notation
$\bW_{R_e}(t(e), h(e)):=\bW^e_{R_e}(s_{t(e)}, s_{h(e)})$.

The partition function \eqref{Ztotal}
is can be re-written as
the partition function of this IRF model:
\beq
Z=\sum_{F\textrm{: face}} W(F) \ .
\label{ZF}
\eeq
Note that we have also included
the Boltzmann weight for self-interactions $\bS^v$
into the definition of the face weight \eqref{Wdef}.
In \eqref{Wdef} we have included a factor of
\beq
\sqrt{\frac{\bW_{2-\gamma-\delta}(d,c) \bW_{2-\beta-\gamma}(b,c)}
{\bW_{2-\alpha-\beta}(a,b)\bW_{2-\alpha-\delta}(a,d)}}
\label{extraW}
\eeq
for later convenience.
This factor cancels out in the definition of the partition 
function \eqref{ZF}, when we sum over faces of the IRF model.
Note that we have
$W_{R_e}(a,b)=W_{2-R_e}(b,a)^{-1}$
thanks to the relation \eqref{gammainv}.

\subsection{Integrability}\label{sec.lattice}

We now claim that the model discussed above 
is integrable.
More precisely the model satisfies the 
star-star relation and consequently the YBE.
This is the main result of the paper.

\paragraph{YBE}

\begin{figure}[htbp]
\centering{\includegraphics[scale=0.25]{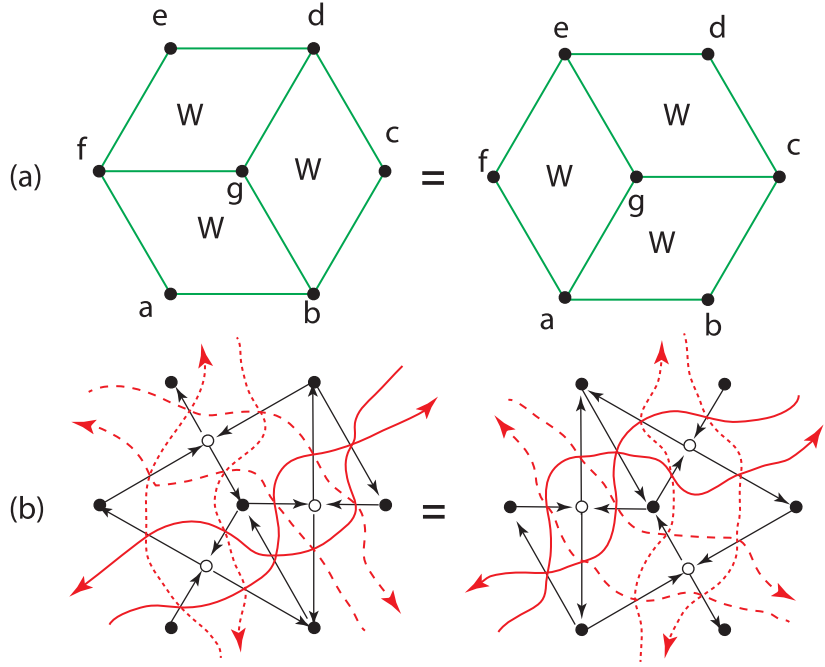}}
\caption{YBE for the IRF model (a) and for the rapidity lines (b).}
\label{fig.YBE}
\end{figure}

The integrability is defined by the famous Yang-Baxter equation.
For the IRF model, the YBE is written as \cite{Baxter:1982zz}
\begin{equation}
\begin{split}
\sum_g \matW{\alpha}{\beta}{\eta}{\zeta}{a}{b}{g}{f} \matW{\zeta}{\eta}{\delta}{\epsilon}{f}{g}{d}{e}
\matW{\eta}{\beta}{\gamma}{\delta}{g}{b}{c}{d}
=\sum_g \matW{\zeta}{\alpha}{\eta}{\epsilon}{f}{a}{g}{e}
 \matW{\alpha}{\beta}{\gamma}{\eta}{a}{b}{c}{g} 
\matW{\eta}{\gamma}{\delta}{\epsilon}{g}{c}{d}{e} \
 , 
\end{split}
\end{equation}
which can be graphically represented as in 
Figure \ref{fig.YBE}(a).
The YBE implies the commutativity of the row-to-row transfer 
matrices, leading to an infinite number of conserved charges \cite{Baxter:1982zz}.

The YBE can also be written in terms of the 
original Boltzmann weights $\bS^v, \bW^e$
(recall the relation \eqref{Wdef}),
which gives the relation in Figure \ref{fig.YBE}(b).
Most of the contributions from \eqref{extraW}
cancel out, however some of them remain and appear in 
Figure \ref{fig.YBE}(b) as extra arrows
connecting two black vertices.

\paragraph{Star-Star Relation}

\begin{figure}[htbp]
\centering{\includegraphics[scale=0.3]{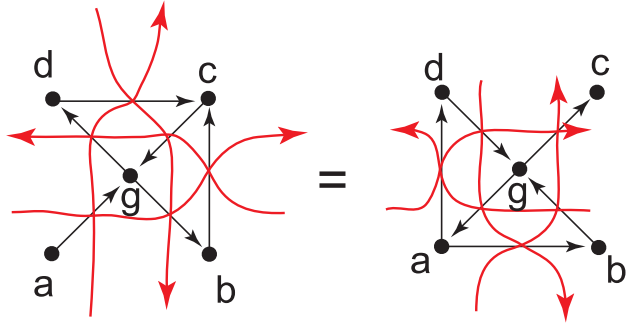}}
\caption{The graphical representation of the star-star relation.}
\label{fig.starstar}
\end{figure}

\begin{figure}[htbp]
\centering{\includegraphics[scale=0.25]{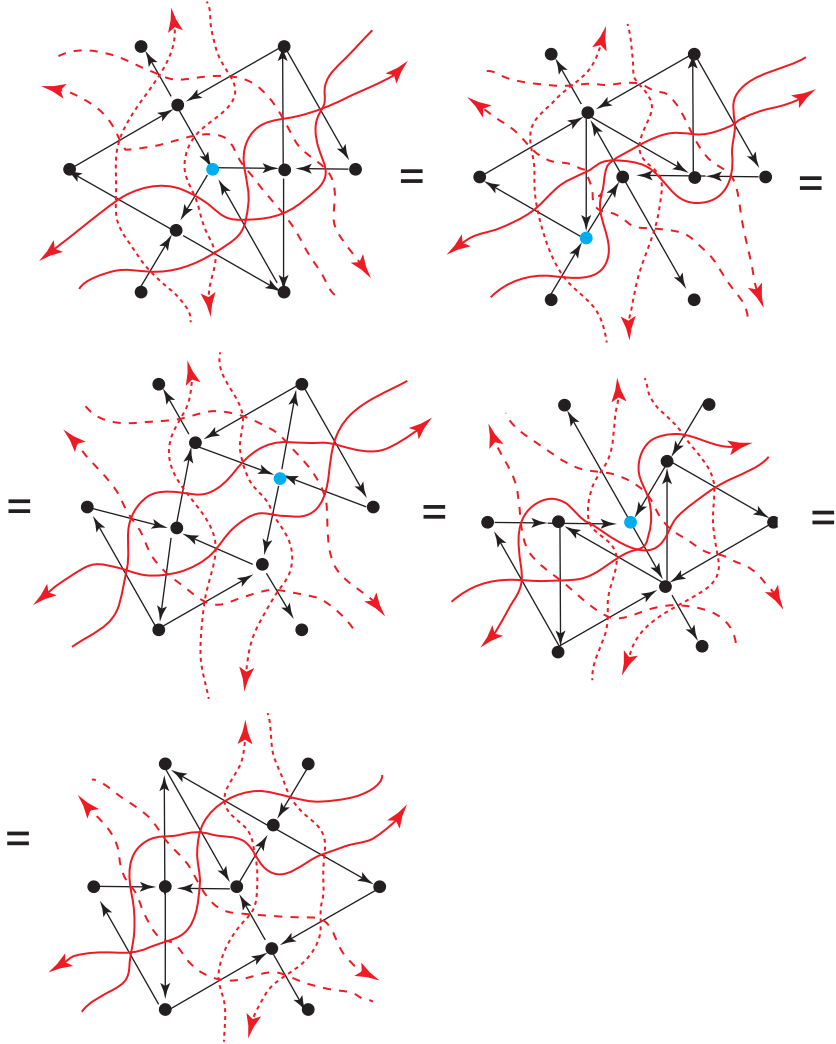}}
\caption{YBE (Figure \ref{fig.YBE}) follows from a sequence of star-star
 relations (Figure \ref{fig.starstar}) \cite{Baxter:1997tn}:
in each step the star-star relation is taken around the blue vertex.}
\label{fig.YBEfromStar}
\end{figure}

Instead of solving the YBE directly,
we solve the the star-star (also called star-to-reversed-star) relation
\cite{Baxter:1986,Bazhanov:1992jqa}:\footnote{
In the definition of the IRF model,
we can choose to integrate out black
vertices instead of white vertices.
This leads to a different IRF model with a face weight
\beq
\overline{W}(F)=\overline{W} 
\left(
\begin{array}{cc}
 d & c \\
 a & b \\
\end{array}
\right)
= 
\sqrt{\frac{\bW_{2-\alpha-\beta}(a,b)\bW_{2-\alpha-\delta}(a,d)}{\bW_{2-\gamma-\delta}(d,c) \bW_{2-\beta-\gamma}(b,c)}
}
\sqrt{\bS^a \bS^c}\, \sum_g 
\bS^g \, 
\bW_{\gamma}(g,a) \bW_{\delta}(b,g) \bW_{\alpha}(g,c)
\bW_{\beta}(d,g) \ .
\eeq
The partition function is again given by \eqref{ZF},
where this time the sum is over the faces of another lattice with 
white vertices only. 
The star-star relation \eqref{starstar}
ensures that the two Boltzmann weights
$W$ and $\overline{W}$
are the same.
}
\begin{equation}
\begin{split}
&\bW_{2-\gamma-\delta}(d,c) \bW_{2-\beta-\gamma}(b,c)
\sum_g \bS^g\, \bW_{\alpha}(a,g) \bW_{\beta}(g,b) 
\bW_{\gamma}(c,g) \bW_{\delta}(g,d) \\
&\quad= \bW_{2-\alpha-\beta}(a,b) \bW_{2-\alpha-\delta}(a,d) 
\sum_g \bS^g\, \bW_{\gamma}(g,a) \bW_{\delta}(b,g)
\bW_{\alpha}(g,c) \bW_{\beta}(d,g) \ .
\end{split}
\label{starstar}
\end{equation}
The graphical representations of this relation is given in Figure \ref{fig.starstar}.\footnote{This 
was called the double Yang-Baxter move in \cite{Yamazaki:2012cp};
it was called the ``double'' since the Yang-Baxter equation is 
often associated with a crossing of three lines, and the star-star relation involves 
two such moves and four lines.
Note however the YBE of our statistical mechanical model can be represented 
as re-shuffling to three sets of parallel lines (Figure \ref{fig.YBE}),
totally six lines.
In integrable model language our statistical mechanical model satisfies the star-star relation,
but not the star-triangle relation. 
}
We can directly verify that the YBE follows from the star-star
relation, see Figure \ref{fig.YBEfromStar}.

It would be desirable to mathematically prove the star-star relation.
For the case with $r=1$ and general $N$, the star-star relation follows from the results of 
\cite{RainsTransf}, as first pointed out by \cite{Dolan:2008qi} 
(see also \cite{Romelsberger:2007ec} for earlier work).\footnote{For the $r=1, N=2$ 
case (where the quiver gauge theory is non-chiral),
the star-star relation follows from the
simpler relation, the star-triangle relation,
which can be proven \cite{Spiridonov:2010em} from the 
elliptic beta integral of \cite{SpiridonovBeta}.}
In this paper the star-star relation will be derived from the 
considerations in 
supersymmetric gauge theory. 
This is be the topic of the next section.

\section{Gauge Theory Origin}\label{sec.quiver}

Let us now comment
on the gauge-theoretic origin of our 
statistical mechanical model.
Interestingly, each of the ingredients introduced in the previous section
has a precise counterpart in 
4d $\scN=1$ supersymmetric gauge theories
(see Table \ref{tab.summary} in section \ref{sec.intro}).
The gauge theory moreover gives generalization of the 
statistical mechanical model to spin lattices
more general than the square lattices of the previous section.
The discussion here will be brief since
most of the ingredients here have been 
discussed in detail in 
\cite{Yamazaki:2012cp,Terashima:2012cx,Xie:2012mr}.

\subsection{Star-Star Relation as Seiberg Duality}

The first step is to define a 4d $\scN=1$ theory associated with the 
spin lattice. 

The idea is to regard the spin lattice as 
the quiver diagram.  Namely we associate an 
$SU(N)$ gauge group to a vertex $v\in V$,
a bifundamental field to an edge $e\in E$, 
and a superpotential term (a trace of the product of bifundamentals)
to a face of the spin lattice.
Note that the orientation of the graph corresponds to the chirality of bifundamental 
multiplets.

It is believed in the literature that the resulting model flows in the IR
to a non-trivial strongly-coupled fixed point. We assume in this paper that this is the
case.\footnote{For periodic boundary 
conditions there is a strong support for this
from the AdS/CFT correspondence (see \cite{Kennaway:2007tq,Yamazaki:2008bt}
and references therein).
See also \cite{Heckman:2012jh}
for the evidence 
of non-trivial IR fixed points for the theories of \cite{Xie:2012mr}.
}
Since the theory has many $U(1)$ global symmetries
the $U(1)$ R-symmetry in the IR can be a mixture of UV 
$U(1)$ R-symmetry with other global $U(1)$ symmetries,
leading to an ambiguity of 
the choice of the R-charge $R_e$
for a bifundamental at an edge $e\in E$.
The ambiguities of the IR R-charges coming from the mixing with 
global symmetries
can be parametrized as in \eqref{difference} \cite{Hanany:2005ss}\footnote{In 
\cite{Hanany:2005ss} the rapidity parameter was identified with the
values of the IR R-charge in the superconformal algebra
determined from $a$-maximization.
However for our purposes we do not adopt this interpretation,
and rather it is crucial to use allow 
the rapidity values to be away from their IR values.
}, where $R_i$ represents the global $U(1)$ symmetry
associated with a rapidity line.
The conditions 
\eqref{sumR} on the R-charges 
represent the vanishing of the beta-functions
for the gauge couplings and superpotential couplings, respectively.

Now
the crucial observation (which goes back to \cite{Spiridonov:2010em}) is that the star-star relation \eqref{starstar}, when translated into the language of 
quiver gauge theories, is precisely the 4d $\scN=1$ Seiberg duality \cite{Seiberg:1994pq}--- in Figure \ref{fig.starstar}
we have Seiberg duality for $N_f=2 N$, where the global symmetry
$SU(2N)$ is broken to $SU(N)\times SU(N)$ by a 
superpotential.\footnote{Mathematically the change of 
quiver in Figure \ref{fig.starstar}
is an example of a quiver mutation.
}
The invariance of the partition function under star-star 
relation is then translated into the invariance of 
some quantity under 4d Seiberg duality.

There is one nice quantity which is indeed invariant under the
Seiberg duality: the 4d lens index, the twisted partition function on 
$S^1\times S^3/\bZ_n$. Here $S^3/\bZ_n$ is the lens space,
where $\bZ_n$ acts on the Hopf fiber of $S^3$.
Let us now discuss this quantity.

\subsection{Partition Function from 4d Lens Index}

The 4d lens index \cite{Benini:2011nc} is defined by
\beq
\scI(p,q,u_i)=\textrm{Tr}_{\, \scH_{S^3/\bZ_r}} \left[
(-1)^F p^{\frac{E+j_2}{3}+j_1} q^{\frac{E+j_2}{3}-j_1}
\right] \ ,
\label{Idef}
\eeq
Here $\scH_{S^3/\bZ_r}$ is the Hilbert space on 
$\scH_{S^3/\bZ_r}$,
$j_1, j_2$ are the generators of $U(1)\times U(1)\in SU(2)\times
SU(2)\sim SO(4)$ and $E$ the generator of translation along $S^1$.
The fugacities for the global symmetries, which can be included
in the definition of \eqref{Idef}, is traded for
the choice of R-charges.
In other words the fugacity for the $i$-th global symmetry is 
given by $(pq)^{R_i}$.

The lens index is one of the most powerful quantities 
known in the literature for 
quantitative analysis of
4d $\scN\ge 1$ supersymmetric gauge theories.
It is more powerful than
their $r=1$ counterparts (see \cite{Razamat:2013jxa,RazamatWillett} 
for examples), and 
also reproduce many partition functions
in lower dimensions, such as the 
3d lens space partition function, the 
3d superconformal index and the 2d sphere partition function
(\cite{Yamazaki:2013fva} and references therein).

Since the 4d lens index is an index, it is invariant under continuous deformations and 
is invariant under the RG flow, as long as we take into account the change of the R-charge
under the RG flow ({\it cf.}\ \cite{Romelsberger:2007ec}).\footnote{In principle the same logic 
applies to any 4d index invariant under Seiberg duality,
however at present the 4d lens index 
is most general index known in the literature.}
This means that
the index can be evaluated in the free field limit.
The resulting expression coincides with
expression found in section \ref{sec.model}.
For example,
the spin variables $s_v=(z_v, m_v)$ are the Wilson lines
for the $SU(N)$ gauge group at vertex $v\in V$.
Since $\pi_1(S^1\times S^3/\bZ_r)=\bZ\times \bZ_r$,
we have continuous Wilson lines
$z_v=(z_{v,1}, \ldots, z_{v,N})\in U(1)^N$, 
(with $\prod_i z_{v,i}=1$)
along the $S^1$, 
or discrete Wilson lines
$m_v=(m_{v,1}, \ldots, m_{v, N})\in (\bZ_r)^{N}$
(with $\sum_i m_{v,i}=0$ modulo $r$)
in the Hopf fiber direction of the manifold.
This explains our choice of spin variables.

The Boltzmann weights $\bS^v, \bW^e$
are the 1-loop determinants for 
vector and chiral multiplets
(including the zero-point contributions 
$\mathcal{S}^v_0, \mathcal{W}^e_0$),
whose explicit expression can
be found in \cite{Benini:2011nc,Yamazaki:2013fva}.\footnote{For 4d $\scN=1$ theories
there was a sign error in version 2 of \cite{Benini:2011nc}, which is corrected in the current paper.
The author would like to thank S.S. Razamat and B. Willet for pointing this out.
He would also like to thank S.S. Razamat for discussion on zero-point contributions
of the 4d lens index.
}
The chiral multiplet 1-loop determinant $\bW^e$ 
corresponds to a nearest-neighbor interaction on the spin lattice
since a bifundamental is charged only with respect to 
two $SU(N)$ gauge groups.

The star-star relation represents the Seiberg duality.\footnote{See 
\cite{RazamatWillett} for checks of Seiberg duality for 4d lens indices.
}
The factors  
\beq
\bW_{2-\gamma-\delta}(d,c), \quad \bW_{2-\beta-\gamma}(b,c),\quad
\bW_{2-\alpha-\beta}(a,b), \quad \bW_{2-\alpha-\delta}(a,d)
\label{meson}
\eeq
in \eqref{starstar}, which can be traced back to \eqref{extraW},
represents the contributions from the mesons in the Seiberg duality.
Since a meson $M$ couples to a quark $q$ and anti-quark $\tilde{q}$
by a superpotential term $W=\tilde{q}Mq$, 
and hence their R-charges should sum up to $2$.
This is represented by the parametrizations of the rapidity parameters
in \eqref{meson} (in particular the numbers ``2'' there).

The statistical model of section \ref{sec.lattice}
uses the square lattice (with periodic boundary conditions).
However the gauge theory considerations apply to more general lattices
determined from the rapidity lines,
as long as the quiver defined from them is connected
(more precisely this is the 
the admissibility conditions of \cite{Ueda:2006jn,Yamazaki:2012cp}).
The vertex of the lattice can be
of $2n$-valent ($n\ge 2$).
The discussion of the star-star relation or equivalently the Seiberg
duality 
works in exactly the same manner.

\section{Further Comments}\label{sec.comments}

Several comments are now in order.

\paragraph{Specialization and Limits}

Since the star-star relation is an equality,
the limits and specializations of the model is still guaranteed to be
integrable.
We can for example take the fugacities $p,q$
to be at root of unity, or regard them as temperatures
and take the high-temperature limit.

We can for example take $r=1$.
The discrete spins $m_v$
are then frozen to be trivial values $m_v=0$, and 
the only non-trivial
degrees of freedom are continuous $U(1)$
spins $z_{v,1}, \ldots, z_{v,N}$. 
As discussed already in introduction, this 
coincides with the Bazhanov-Sergeev model \cite{Bazhanov:2011mz}
(up to the normalization constant of the partition function
which is not essential for the discussion of integrability here).
This model is known to reproduce a number of
known integrable models \cite{Bazhanov:2007mh,Bazhanov:2010kz,Bazhanov:2011mz}, such as
the Faddeev-Volkov model \cite{Volkov:1992uv,FaddeevVolkovAbelian}\footnote{The reduction 
from the Bazhanov-Sergeev model to 
Faddeev-Volkov model 
corresponds to 
the dimensional reduction and Higgsing of
the 4d $\scN=1$ theory down to 3d $\scN=2$ theory.
In this process 2d spin system lifts to the geometry of 
a hyperbolic 3-manifold \cite{Yamazaki:2012cp,Terashima:2012cx}.
}, the chiral Potts model \cite{AuYang:1987zc,Baxter:1987eq} and the
 Kashiwara-Miwa model \cite{Kashiwara:1986tu}.
The last two models are different generalizations of the 
Fateev-Zamolodchikov model \cite{Fateev:1982wi}.

Interestingly,
many of these models have discontinuous spins:
the values of the continuous spins in the Bazhanov-Sergeev model 
are frozen to discrete values--- this happens, for example, when 
we choose the temperature-like parameters $p,q$
to be at root of unity.
It is interesting to ask whether the inverse procedure is possible,
namely whether we can construct a model with all spins continuous 
such that our model is reproduced as a high-temperature limit.
Such a model, if it exists, should be more powerful than the model discussed in
this paper and could have some gauge theory implications, for example
as an index of some 5d gauge theory.

\paragraph{Further Generalizations}

There are several natural generalizations of our model.
For example, we can take the 
rank of gauge groups
(and hence the number of components of spins)
to be $v$-dependent: the gauge group at $v$ is $SU(N_v)$.
It is straightforward to write down the corresponding statistical model.
However, not all the possible values of $N_v$ are allowed
since we have the ensure the absence of anomalies
and the existence of the non-trivial IR fixed point.
Moreover the ranks of the gauge groups will change under the Seiberg
duality, and for this reason the corresponding YBE is more general than the one 
typically considered in integrable models.
The same subtlety happens when we consider
$S^1\times S^2$ index of 3d $\scN=2$ quiver Chern-Simons-matter theories
and the Giveon-Kutasov dualities \cite{Giveon:2008zn} among them.

\paragraph{Comparison with Gauge/Bethe Correspondence}

The Gauge/YBE correspondence \eqref{GaugeYBE} discussed in this paper
is reminiscent of a 
similar correspondence in the literature, 
the Gauge/Bethe correspondence \cite{Nekrasov:2009uh,Nekrasov:2009ui}:
\beq
\scW_{\rm eff}(\sigma)=Y(\sigma) \ .
\label{GaugeBethe}
\eeq
On the left hand side of \eqref{GaugeBethe}
we have the effective twisted
super potential $\scW_{\rm eff}(\sigma)$
of a 2d $\scN=(2,2)$ theory as a function of 
the scalar in the twisted multiplet $\Sigma=\overline{D}_+ D_- V$.
On the right hand side we have the Yang-Yang function \cite{Yang:1968rm} 
$Y(\sigma)$. The derivative of this function gives the Bethe Ansatz equation,
which coincides with the vacuum equation of the 
2d $\scN=(2,2)$ theory:
\beq
\exp\left(\frac{\partial \scW_{\rm eff}}{\partial \sigma} \right)=
\exp\left(\frac{\partial Y}{\partial \sigma}\right)=1\ .
\eeq

While the two relations \eqref{GaugeYBE}, \eqref{GaugeBethe}
are similar, there are crucial differences between the
two.\footnote{Despite 
the differences it would be interesting to ask if the two relations
are related more directly, at least for a special class of theories.
In fact, if we choose $r=1$
the dimensional reduction of the 4d index
in the Gauge/YBE correspondence
gives the 3d ellipsoid partition function $S^3_b$,
and then the 2d effective twisted superpotential in the 
$b\to 0$ limit. This reduction preserves integrability
of the associated statistical mechanical model.
Since we have 2d $\scN=(2,2)$ theory,
we could then consider the
corresponding integrable model of the 2d theory through the Gauge/Bethe correspondence, if it exists at all.}
One crucial difference is that 
in Gauge/YBE we have the partition function
of an integrable model,
whereas in Gauge/Bethe we have
the Yang-Yang function.
This is a huge difference when we want to 
find {\it new} integrable models from the correspondence.

In Gauge/Bethe correspondence 
it is in general
rather hard to identify the corresponding integrable model for 
a given 2d $(2,2)$ theory---there is no general algorithm 
to recover the R-matrices from
the Bethe Ansatz equation.
In fact, it is not known in general whether such an integrable model
really exists for a given 2d $\scN=(2,2)$ theory.\footnote{The
exception 
is the case where the 2d $(2,2)$ theory comes from 4d $\scN=2$ 
class $\mathcal{S}$ theories
on the Omega-background \cite{Nekrasov:2002qd} with 
equivariant parameters $\epsilon_1=0, \epsilon_2\ne 0$ \cite{Nekrasov:2009rc}:
the corresponding integrable model in this case
is Hitchin integrable model.
However, this is not really a new integrable model 
found from the correspondence,
and moreover the story does not generalize to more general 2d $\scN=(2,2)$ theories.}

This should be contrasted with the Gauge/YBE correspondence 
discussed in this paper.
There the gauge theory answer directly gives 
precisely the R-matrix and the partition function of 
the integrable model, and the 
the reason {\it why} the model is integrable 
has a direct explanation from the 4d Seiberg duality.
This direct relation was the reason that in this paper
we could construct new integrable models
from the correspondence.

\paragraph{Underlying Algebra}

One of the systematic methods to construct solutions to YBE is to 
define the R-matrix as an intertwiner of the tensor product of representations.
The relevant algebra for the $r=1$ case is the Sklyanin algebra \cite{Sklyanin:1982tf}
and their generalizations \cite{CherednikGeneralized} 
(see e.g.\ \cite{Zabrodin:2010qm}), which are 2-parameter 
deformations of the universal enveloping algebra $U(\mathfrak{gl}_N)$.
Such an algebra acts on the operators of {\it a class of} supersymmetric gauge theories,
and plays the role similar to that of the Yangian in the Gauge/Bethe correspondence \cite{Nekrasov:2009uh,Nekrasov:2009ui}.

\paragraph{Knot Invariants}

It would be interesting to study the knots invariants 
associated with our solutions to the YBE ({\it cf.}\ \cite{Wadati:1989ud}),
see \cite{Wu2} for related discussion for the chiral Potts model.

\section{Concluding Remarks}\label{sec.conclusion}

We proposed a 2d statistical 
spin lattice model with 
nearest-neighbor and self interactions.
At each vertex we have a hybrid of continuous spins and discrete spins,
each of which has $N$ components.
The Boltzmann weights are given in \eqref{Wweight}, \eqref{Sweight}
and are written in terms of lens elliptic gamma functions \eqref{egamma1}. 
The model can be reformulated either as a vertex model
or an IRF model.

The model is integrable (satisfy the star-star relation and hence the YBE)
if we assume 
the invariance of the 4d lens index under the Seiberg duality.
Alternatively, the mathematical proof of integrability 
can be thought of as another non-trivial test of 
the existence of non-trivial IR fixed points
and the 4d Seiberg duality.

It is important to mathematically prove the
integrability of the model. We can also study 
various properties of the model, such as free energies and correlation
functions, and their gauge theory 
interpretations.\footnote{Thermodynamics limit of the model 
could have direct geometric
interpretation ({\it cf.}\ \cite[section 5]{Yamazaki:2012cp}),
perhaps along the lines of \cite{KenyonOkounkovSheffield,Ooguri:2009ri}.
}

\bigskip

In this paper
we have concentrated on the application of supersymmetric
gauge theories to integrable models.
We hope that our results are of interest to
experts on integrable models.
While this is interesting in its own right, 
experts on supersymmetric gauge theorists might 
be interested in a different question,
namely if our results sheds any new light on 
our understanding of 
4d $\scN=1$ supersymmetric gauge theories.

In this respect, one fundamental question is whether
the integrable structure discussed in this paper goes
beyond the 4d lens index, and extends to the 
full physics of the IR fixed points.

We do not have the complete answer yet, however
let us here point out that
we have already identified dualities of quiver gauge theories 
corresponding to YBE: it is the sequence of four Seiberg dualities,
and the quiver diagrams for the two theories
are shown in Figure \ref{fig.YBE}.
This duality among supersymmetric gauge theories
requires several $SU(N)$
gauge groups, which is probably the reason why 
it has not been paid much attention so far.

Recall that the R-matrix in integrable models
is the linear map $R: V\otimes V \to V\otimes V$.
This corresponds to a 4d $\scN=1$ theory 
$\mathcal{T}[R]$, whose quiver diagram is a simply a ``star''
of Figure \ref{fig.faceweight}---
if we un-gauge the gauge groups at the four black vertices,
the theory is has global symmetry $SU(N)^4$,
and the $SU(N)$ global symmetry plays the role 
similar to that of $V$. 
The counterpart of taking a product of R-matrices
is to concatenate two quiver diagrams, 
by re-gauging the diagonal part of global $SU(N)\times SU(N)$
symmetries of the theories associated with the quivers.
Finally, the counterpart of YBE is 
the following duality between two 4d $\scN=1$
theories:
\beq
\mathcal{T}[R_{12}] \circ \mathcal{T}[R_{13}] \circ \mathcal{T}[R_{23}]  
\quad 
\textrm{is dual to}
\quad \mathcal{T}[R_{23}] \circ \mathcal{T}[R_{13}] \circ
\mathcal{T}[R_{12}] \ ,
\label{catYBE}
\eeq
where the symbol $\circ$
represents the gauging the diagonal subgroup of 
product global symmetries of two theories.\footnote{More precisely 
we need to include two extra bifundamentals chiral
multiplets (mesons)
to both sides
of \eqref{catYBE} (Figure \ref{fig.YBE}).
}
The duality \eqref{catYBE}, which we call the {\it Yang-Baxter duality},
is more powerful than the equivalence of 4d lens indices.
For example, the physical Hilbert spaces of the two theories should
coincide in the IR, which can be thought of 
as a version of categorification of the Gauge/YBE correspondence.

The identification of the YBE with the 
duality \eqref{catYBE}
makes it clear that for the discussion of 
integrability we need to consider
not a single theory, but a class of them related by
gauging/un-gauging procedures:
in the spirit of \cite{Yamazaki:2013xva},
what is relevant here is ``integrability in theory space''.

It would be interesting to analyze the duality \eqref{catYBE}
further. The fact that the YBE is more general than the
star-star relation in integrable model suggests that
there should be generalizations of theory $\mathcal{T}[R]$
and dualities among them, 
which cannot be decomposed into Seiberg dualities.
Such dualities will give an ultimate reason
{\it why} gauge-theoretic quantities have integrable structures.

\section*{Acknowledgments}

The author would like to take this opportunity to 
thank M. Jimbo for his excellent textbook ``quantum groups and Yang-Baxter equations'' (Springer, 1990, in Japanese).
The author has been mesmerized by the beauty of integrable models
ever since he read the textbook a decade ago.
The author would also like thank S. S. Razamat and B. Willett
for stimulating discussion on the 4d lens index and for informing the author of
their forthcoming paper \cite{RazamatWillett}.
He would like to thank Yukawa Institute for Theoretical Physics, Kyoto University
(YKIS 2012) for hospitality where part of this work was performed.
The results of this paper was presented at the pre-strings meeting 
``Exact Results in String/M-theory'' (KIAS, June 2013) and
the author would like to thank the audience for feedback.
This research was supported by World Premier
International Research Center Initiative (WPI Initiative), MEXT, Japan.

\appendix

\section{Lens Elliptic Gamma Function}\label{sec.gamma}

In this appendix we briefly summarize the definition and the properties of the lens elliptic gamma function
used in the main text.

Let us choose a positive integer $r$ and 
an element $\mmod{m}$ in $\bZ_r$,
where $\mmod{m}$ represents the $m$ modulo $r$, {\it
 i.e.}\ $0\le \mmod{m} \le r-1$ and $\mmod{m}\equiv m$ modulo $r$.
The lens elliptic gamma function elliptic gamma function $\Gamma_{r,\mmod{m}}(x;p,q)$
is defined by
\beq
\Gamma_{r, \mmod{m}}(x;p,q)=
 \frac{\displaystyle\prod_{n_1, n_2\ge 0,\, n_1-n_2\equiv \mmod{-m}\, (\textrm{mod } r) } \left( 1-x^{-1} p^{n_1+1} q^{n_2+1} \right)}
 {\displaystyle\prod_{n_1, n_2\ge 0,\, n_1-n_2\equiv \mmod{m}\, (\textrm{mod } r) } \left(1-x\, p^{n_1} q^{n_2} \right)} \ .
\label{egamma1}
\eeq
We can rewrite this function as
\beq
\Gamma_{r,\mmod{m}}(x;p,q)=
\Gamma(x p^{\mmod{m}};pq, p^r)\, \Gamma(x q^{r-\mmod{m}};pq, q^r) \ ,
\eeq
where $\Gamma(x;p,q)=\Gamma_{1,0}(x;p,q)$ is the ordinary elliptic 
gamma function found in the literature
\beq
\Gamma(x;p,q)=\prod_{n_1, n_2\ge 0} \frac{1-x^{-1} p^{n_1+1} q^{n_2+1}}{1-x\, p^{n_1} q^{n_2}} \ .
\label{egammaordinary}
\eeq
For $m=0$ the lens elliptic gamma function simplifies to
\beq
\Gamma_{r, 0}(x;p,q)=\Gamma(x;pq, p^r)\, \Gamma(xq^r;pq,q^r)
=\Gamma(x;p^r, q^r) \ .
\label{egamma0}
\eeq
We can easily prove the relations
\beq
\Gamma_{r,\mmod{-m}}\left(\frac{pq}{x};p,q\right)=\Gamma_{r,\mmod{m}}(x;p,q)^{-1} \ ,
\label{gammainv}
\eeq


\bibliographystyle{JHEP}
\bibliography{LensInt}

\def\cprime{$'$}
\providecommand{\href}[2]{#2}\begingroup\raggedright\begin{thebibliography}{10}

\bibitem{Benini:2011nc}
F.~Benini, T.~Nishioka, and M.~Yamazaki, {\it {4d Index to 3d Index and 2d
  TQFT}},  {\em Phys.Rev.} {\bf D86} (2012) 065015,
  [\href{http://xxx.lanl.gov/abs/1109.0283}{{\tt arXiv:1109.0283}}].

\bibitem{Xie:2012mr}
D.~Xie and M.~Yamazaki, {\it {Network and Seiberg Duality}},  {\em JHEP} {\bf
  1209} (2012) 036, [\href{http://xxx.lanl.gov/abs/1207.0811}{{\tt
  arXiv:1207.0811}}].

\bibitem{Franco:2012mm}
S.~Franco, {\it {Bipartite Field Theories: from D-Brane Probes to Scattering
  Amplitudes}},  \href{http://xxx.lanl.gov/abs/1207.0807}{{\tt
  arXiv:1207.0807}}.

\bibitem{Hanany:1997tb}
A.~Hanany and A.~Zaffaroni, {\it {On the realization of chiral four-dimensional
  gauge theories using branes}},  {\em JHEP} {\bf 9805} (1998) 001,
  [\href{http://xxx.lanl.gov/abs/hep-th/9801134}{{\tt hep-th/9801134}}].

\bibitem{Hanany:1998it}
A.~Hanany and A.~M. Uranga, {\it {Brane boxes and branes on singularities}},
  {\em JHEP} {\bf 9805} (1998) 013,
  [\href{http://xxx.lanl.gov/abs/hep-th/9805139}{{\tt hep-th/9805139}}].

\bibitem{Hanany:2005ve}
A.~Hanany and K.~D. Kennaway, {\it {Dimer models and toric diagrams}},
  \href{http://xxx.lanl.gov/abs/hep-th/0503149}{{\tt hep-th/0503149}}.

\bibitem{Yamazaki:2012cp}
M.~Yamazaki, {\it {Quivers, YBE and 3-manifolds}},  {\em JHEP} {\bf 1205}
  (2012) 147, [\href{http://xxx.lanl.gov/abs/1203.5784}{{\tt
  arXiv:1203.5784}}].

\bibitem{Terashima:2012cx}
Y.~Terashima and M.~Yamazaki, {\it {Emergent 3-manifolds from 4d Superconformal
  Indices}},  {\em Phys.Rev.Lett.} {\bf 109} (2012) 091602,
  [\href{http://xxx.lanl.gov/abs/1203.5792}{{\tt arXiv:1203.5792}}].

\bibitem{Kinney:2005ej}
J.~Kinney, J.~M. Maldacena, S.~Minwalla, and S.~Raju, {\it {An Index for 4
  Dimensional Super Conformal Theories}},  {\em Commun. Math. Phys.} {\bf 275}
  (2007) 209--254, [\href{http://xxx.lanl.gov/abs/hep-th/0510251}{{\tt
  hep-th/0510251}}].

\bibitem{Romelsberger:2005eg}
C.~Romelsberger, {\it {Counting chiral primaries in N = 1, d=4 superconformal
  field theories}},  {\em Nucl.Phys.} {\bf B747} (2006) 329--353,
  [\href{http://xxx.lanl.gov/abs/hep-th/0510060}{{\tt hep-th/0510060}}].

\bibitem{Bazhanov:2010kz}
V.~V. Bazhanov and S.~M. Sergeev, {\it {A Master solution of the quantum
  Yang-Baxter equation and classical discrete integrable equations}},
  \href{http://xxx.lanl.gov/abs/1006.0651}{{\tt arXiv:1006.0651}}.

\bibitem{Bazhanov:2011mz}
V.~V. Bazhanov and S.~M. Sergeev, {\it {Elliptic gamma-function and multi-spin
  solutions of the Yang-Baxter equation}},  {\em Nucl.Phys.} {\bf B856} (2012)
  475--496, [\href{http://xxx.lanl.gov/abs/1106.5874}{{\tt arXiv:1106.5874}}].

\bibitem{Bazhanov:2013bh}
V.~V. Bazhanov, A.~P. Kels, and S.~M. Sergeev, {\it {Comment on star-star
  relations in statistical mechanics and elliptic gamma-function identities}},
  \href{http://xxx.lanl.gov/abs/1301.5775}{{\tt arXiv:1301.5775}}.

\bibitem{Spiridonov:2010em}
V.~Spiridonov, {\it {Elliptic beta integrals and solvable models of statistical
  mechanics}},  \href{http://xxx.lanl.gov/abs/1011.3798}{{\tt
  arXiv:1011.3798}}.

\bibitem{Baxter:1986}
R.~Baxter, {\it {The Yang-Baxter Equations and the Zamolodchikov Model}},  {\em
  Physica} {\bf D18} (1986) 321--347.

\bibitem{Bazhanov:1992jqa}
V.~Bazhanov and R.~Baxter, {\it {New solvable lattice models in
  three-dimensions}},  {\em J.Statist.Phys.} {\bf 69} (1992) 453--585.

\bibitem{Baxter:1997tn}
R.~Baxter, {\it {Star-triangle and star-star relations in statistical
  mechanics}},  {\em Int.J.Mod.Phys.} {\bf B11} (1997) 27--37.

\bibitem{Baxter:1978xr}
R.~Baxter, {\it {Solvable eight vertex model on an arbitrary planar lattice}},
  {\em Phil.Trans.Roy.Soc.Lond.} {\bf 289} (1978) 315--346.

\bibitem{Baxter:1982zz}
R.~Baxter, {\em {Exactly solved models in statistical mechanics}}.
\newblock Dover, 2007.

\bibitem{RainsTransf}
E.~M. Rains, {\it Transformations of elliptic hypergeometric integrals},  {\em
  Ann. of Math. (2)} {\bf 171} (2010), no.~1 169--243.

\bibitem{Dolan:2008qi}
F.~Dolan and H.~Osborn, {\it {Applications of the Superconformal Index for
  Protected Operators and q-Hypergeometric Identities to N=1 Dual Theories}},
  {\em Nucl.Phys.} {\bf B818} (2009) 137--178,
  [\href{http://xxx.lanl.gov/abs/0801.4947}{{\tt arXiv:0801.4947}}].

\bibitem{Romelsberger:2007ec}
C.~Romelsberger, {\it {Calculating the Superconformal Index and Seiberg
  Duality}},  \href{http://xxx.lanl.gov/abs/0707.3702}{{\tt arXiv:0707.3702}}.

\bibitem{SpiridonovBeta}
V.~P. Spiridonov, {\it On the elliptic beta function},  {\em Uspekhi Mat. Nauk}
  {\bf 56} (2001), no.~1(337) 181--182.

\bibitem{Kennaway:2007tq}
K.~D. Kennaway, {\it {Brane Tilings}},  {\em Int.J.Mod.Phys.} {\bf A22} (2007)
  2977--3038, [\href{http://xxx.lanl.gov/abs/0706.1660}{{\tt
  arXiv:0706.1660}}].

\bibitem{Yamazaki:2008bt}
M.~Yamazaki, {\it {Brane Tilings and Their Applications}},  {\em Fortsch.Phys.}
  {\bf 56} (2008) 555--686, [\href{http://xxx.lanl.gov/abs/0803.4474}{{\tt
  arXiv:0803.4474}}]. Master's Thesis.

\bibitem{Heckman:2012jh}
J.~J. Heckman, C.~Vafa, D.~Xie, and M.~Yamazaki, {\it {String Theory Origin of
  Bipartite SCFTs}},  \href{http://xxx.lanl.gov/abs/1211.4587}{{\tt
  arXiv:1211.4587}}.

\bibitem{Hanany:2005ss}
A.~Hanany and D.~Vegh, {\it {Quivers, tilings, branes and rhombi}},  {\em JHEP}
  {\bf 0710} (2007) 029, [\href{http://xxx.lanl.gov/abs/hep-th/0511063}{{\tt
  hep-th/0511063}}].

\bibitem{Seiberg:1994pq}
N.~Seiberg, {\it {Electric - magnetic duality in supersymmetric nonAbelian
  gauge theories}},  {\em Nucl.Phys.} {\bf B435} (1995) 129--146,
  [\href{http://xxx.lanl.gov/abs/hep-th/9411149}{{\tt hep-th/9411149}}].

\bibitem{Razamat:2013jxa}
S.~S. Razamat and M.~Yamazaki, {\it {S-duality and the N=2 Lens Space Index}},
  \href{http://xxx.lanl.gov/abs/1306.1543}{{\tt arXiv:1306.1543}}.

\bibitem{RazamatWillett}
S.~S. Razamat and B.~Willett, {\it {Global properties of supersymmetric
  theories and the lens space}},  to appear.

\bibitem{Yamazaki:2013fva}
M.~Yamazaki, {\it {Four-dimensional superconformal index reloaded}},  {\em
  Theor.Math.Phys.} {\bf 174} (2013) 154--166.

\bibitem{Ueda:2006jn}
K.~Ueda and M.~Yamazaki, {\it {A note on dimer models and McKay quivers}},
  {\em Commun.Math.Phys.} {\bf 301} (2011) 723--747,
  [\href{http://xxx.lanl.gov/abs/math/0605780}{{\tt math/0605780}}].

\bibitem{Bazhanov:2007mh}
V.~V. Bazhanov, V.~V. Mangazeev, and S.~M. Sergeev, {\it {Faddeev-Volkov
  solution of the Yang-Baxter equation and discrete conformal symmetry}},  {\em
  Nucl.Phys.} {\bf B784} (2007) 234--258,
  [\href{http://xxx.lanl.gov/abs/hep-th/0703041}{{\tt hep-th/0703041}}].

\bibitem{Volkov:1992uv}
A.~Volkov, {\it {Quantum Volterra model}},  {\em Phys.Lett.} {\bf A167} (1992)
  345--355.

\bibitem{FaddeevVolkovAbelian}
L.~Faddeev and A.~Y. Volkov, {\it Abelian current algebra and the {V}irasoro
  algebra on the lattice},  {\em Phys. Lett. B} {\bf 315} (1993), no.~3-4
  311--318.

\bibitem{AuYang:1987zc}
H.~Au-Yang, B.~M. McCoy, J.~H. Perk, S.~Tang, and M.-L. Yan, {\it {Commuting
  transfer matrices in the chiral Potts models: Solutions of Star triangle
  equations with genus $>$ 1}},  {\em Phys.Lett.} {\bf A123} (1987) 219--223.

\bibitem{Baxter:1987eq}
R.~Baxter, J.~Perk, and H.~Au-Yang, {\it {New solutions of the star triangle
  relations for the chiral Potts model}},  {\em Phys.Lett.} {\bf A128} (1988)
  138--142.

\bibitem{Kashiwara:1986tu}
M.~Kashiwara and T.~Miwa, {\it A class of elliptic solutions to the star
  triangle relation},  {\em Nucl.Phys.} {\bf B275} (1986) 121.

\bibitem{Fateev:1982wi}
V.~Fateev and A.~Zamolodchikov, {\it {Selfdual solutions of the star triangle
  relations in Z(N) models}},  {\em Phys.Lett.} {\bf A92} (1982) 37--39.

\bibitem{Giveon:2008zn}
A.~Giveon and D.~Kutasov, {\it {Seiberg Duality in Chern-Simons Theory}},  {\em
  Nucl.Phys.} {\bf B812} (2009) 1--11,
  [\href{http://xxx.lanl.gov/abs/0808.0360}{{\tt arXiv:0808.0360}}].

\bibitem{Nekrasov:2009uh}
N.~A. Nekrasov and S.~L. Shatashvili, {\it {Supersymmetric vacua and Bethe
  ansatz}},  {\em Nucl.Phys.Proc.Suppl.} {\bf 192-193} (2009) 91--112,
  [\href{http://xxx.lanl.gov/abs/0901.4744}{{\tt arXiv:0901.4744}}].

\bibitem{Nekrasov:2009ui}
N.~A. Nekrasov and S.~L. Shatashvili, {\it {Quantum integrability and
  supersymmetric vacua}},  {\em Prog.Theor.Phys.Suppl.} {\bf 177} (2009)
  105--119, [\href{http://xxx.lanl.gov/abs/0901.4748}{{\tt arXiv:0901.4748}}].

\bibitem{Yang:1968rm}
C.-N. Yang and C.~Yang, {\it {Thermodynamics of one-dimensional system of
  bosons with repulsive delta function interaction}},  {\em J.Math.Phys.} {\bf
  10} (1969) 1115--1122.

\bibitem{Nekrasov:2002qd}
N.~A. Nekrasov, {\it {Seiberg-Witten Prepotential From Instanton Counting}},
  {\em Adv. Theor. Math. Phys.} {\bf 7} (2004) 831--864,
  [\href{http://xxx.lanl.gov/abs/hep-th/0206161}{{\tt hep-th/0206161}}].

\bibitem{Nekrasov:2009rc}
N.~A. Nekrasov and S.~L. Shatashvili, {\it {Quantization of Integrable Systems
  and Four Dimensional Gauge Theories}},
  \href{http://xxx.lanl.gov/abs/0908.4052}{{\tt arXiv:0908.4052}}.

\bibitem{Sklyanin:1982tf}
E.~Sklyanin, {\it {Some algebraic structures connected with the Yang-Baxter
  equation}},  {\em Funct.Anal.Appl.} {\bf 16} (1982) 263--270.

\bibitem{CherednikGeneralized}
I.~V. Cherednik, {\it Some finite-dimensional representations of generalized
  {S}klyanin algebras},  {\em Funktsional. Anal. i Prilozhen.} {\bf 19} (1985),
  no.~1 89--90.

\bibitem{Zabrodin:2010qm}
A.~Zabrodin, {\it {Intertwining operators for Sklyanin algebra and elliptic
  hypergeometric series}},  {\em J.Geom.Phys.} {\bf 61} (2011) 1733--1754,
  [\href{http://xxx.lanl.gov/abs/1012.1228}{{\tt arXiv:1012.1228}}].

\bibitem{Wadati:1989ud}
M.~Wadati, T.~Deguchi, and Y.~Akutsu, {\it {Exactly Solvable Models and Knot
  Theory}},  {\em Phys.Rept.} {\bf 180} (1989) 247.

\bibitem{Wu2}
F.~Y. Wu, P.~Pant, and C.~King, {\it The chiral {P}otts model and its
  associated link invariant},  {\em J. Statist. Phys.} {\bf 78} (1995), no.~5-6
  1253--1276.

\bibitem{KenyonOkounkovSheffield}
R.~Kenyon, A.~Okounkov, and S.~Sheffield, {\it Dimers and amoebae},  {\em Ann.
  of Math. (2)} {\bf 163} (2006), no.~3 1019--1056.

\bibitem{Ooguri:2009ri}
H.~Ooguri and M.~Yamazaki, {\it {Emergent Calabi-Yau Geometry}},  {\em
  Phys.Rev.Lett.} {\bf 102} (2009) 161601,
  [\href{http://xxx.lanl.gov/abs/0902.3996}{{\tt arXiv:0902.3996}}].

\bibitem{Yamazaki:2013xva}
M.~Yamazaki, {\it {Entanglement in Theory Space}},
  \href{http://xxx.lanl.gov/abs/1304.0762}{{\tt arXiv:1304.0762}}.

\end{thebibliography}\endgroup
\end{document}